# Photo-induced bond breaking during phase separation kinetics of block copolymer melts: A dissipative particle dynamics study

Ashish Kumar Singh[1], Avinash Chauhan[1], Sanjay Puri[2*], and Awaneesh Singh[1*]

[1]Department of Physics, Indian Institute of Technology (BHU), Varanasi-221005, India.

[2]School of Physical Sciences, Jawaharlal Nehru University, New Delhi-110067, India.

*Emails: awaneesh.phy@iitbhu.ac.in; purijnu@gmail.com

## Abstract

Using dissipative particle dynamics (DPD) simulation method, we study the phase separation dynamics in block copolymer (BCP) melt in $d = 3$, subjected to external stimuli such as light. An initial homogeneous BCP melt is rapidly quenched to a temperature $T < T_c$, where $T_c$ is the critical temperature. We then let the system go through alternate light "on" and "off" cycles. An on-cycle breaks the stimuli-sensitive bonds connecting both the blocks A and B in BCP melt, and during the off-cycle, broken bonds reconnect. By simulating the effect of light, we isolate scenarios where phase separation begins with the light off (set 1); the cooperative interactions within the system allow it to undergo microphase separation. When the phase separation starts with the light on (set 2), the system undergoes macrophase separation due to the bond breaking. Here, we report the role of alternate cycles on domain morphology by varying bond-breaking probability for both the sets 1 and 2, respectively. We observe that the scaling functions depend upon the conditions mentioned above that change the time scale of the evolving morphologies in various cycles. However, in all the cases, the average domain size respects the power-law growth: $R(t) \sim t^\phi$ at late times, here $\phi$ is the dynamic growth exponent. After a short-lived diffusive growth ($\phi \sim 1/3$) at early times, $\phi$ illustrates a crossover from the viscous hydrodynamic ($\phi \sim 1$) to the inertial hydrodynamic ($\phi \sim 2/3$) regimes at late times.

# 1. Introduction

In recent years, there has been a significant increase of attention in using light as a smart stimulus for externally triggered altering and switching the properties of soft materials.[1–10] The fabrication of soft materials using photoinitiated free radical polymerization (FRP) is unambiguously a robust and versatile technique.[5–14] This technique controls the design and application of the next-generation responsive soft materials with several advanced applications in biomaterials, adhesives, coatings, self-healing, and add remarkable innovations in additive manufacturing of 3D-printing systems.[7–10] Often, these systems utilize photo-stimulated bond breaking to facilitate the exchange of broken polymer chains and complement further with polymerization and crosslinking in the light "on" state, which is an otherwise covalently bonded during the "off" state.[7–10]

Many recent studies have exhibited the phase separation induced by chemical reaction where resulting morphologies, typically known as the modulated phase-pattern, are the consequence of the competition between two opposing phenomena: phase separation and chemical reaction. The chemical reactions usually include either polymerization or crosslinking reaction or both, or simple reactions such as $A \rightleftharpoons B$.[15–20] The unfavorable competition between phase separation and chemical reaction yields a variety of morphologies, depending upon their strength and the spatial distribution.[18,19,21,22] The chemical reactions involved in these works are usually thermally activated. Therefore, it would be fascinating to use light to efficiently control the kinetics of phase separation that remains mostly unexplored till date where photo-reactions (*e.g.*, initiation or termination) can be executed independently from the temperature and pressure of the system.[17]

In particular, the phase separation kinetics is well studied in diverse systems such as multicomponent fluids, polymeric solutions, metallic alloys, and glasses, etc.[23–25] A rapid temperature quench of the homogeneous mixture inside the miscibility gap makes the system thermodynamically unstable. The small instabilities in local density field grow and coarsen into domains of various phases.[26–29] Ample interest was put into the kinetics of separation of immiscible phases of block copolymer (BCP) melts due to its technological importance and, thus, comprehensively studied via experiments and simulations for years.[30–34] Usually, the phase separation process is assisted by the enthalpy of mixing of incompatible blocks coupled with entropy that favors the blending. A few essential parameters that regulate the domain morphologies such as lamellae, gyroids, cylinders, and spheres are the volume fraction of the blocks, the degree of polymerization, and the Flory-Huggins interaction parameter.[33–36] Other external parameters, such as electrical or mechanical fields, may also control the self-assembly of BCP melt.[37,38]

It is now well-understood that domain coarsening is a scaling phenomenon[26,27]. The growth of domains, particularly for the pure and isotropic $3d$ systems, can be characterized by the power-law: $R(t) \sim t^\phi$; here, $R(t)$ is the characteristic length scale, and $\phi$ is the growth exponent that illustrates the dominant transport mechanism driving the phase segregation.[27] For the spinodal decomposition of macrophase separating fluids[27], three distinct growth regimes were seen at successive time scales:[39–41]

$$R(t) \sim \begin{cases} (D\sigma t)^{1/3}, & R(t) \ll (D\eta)^{1/2}, & \text{(diffusive)} \\ \sigma t/\eta, & (D\eta)^{1/2} \ll R(t) \ll \eta^2/\rho\sigma, & \text{(viscous hydrodynamics)} \\ (\sigma t^2/\rho)^{1/3}, & \eta^2/\rho\sigma \ll R(t), & \text{(inertial hydrodynamics)} \end{cases} \quad (1)$$

where $D$ implies the diffusion coefficient; $\eta$ is the viscosity; $\rho$ is the density, and $\sigma$ represents the surface tension. A few experimental reports show the evidence for a crossover from the diffusive to the viscous hydrodynamic regime; however, the inertial growth regime has not been observed by any experiment till date.[42–44] Nonetheless, recently, a crossover from the viscous ($\phi \sim 1$) to the inertial hydrodynamic regime ($\phi \sim 2/3$) is reported for phase separating binary/ternary fluids in $d = 3$, utilizing dissipative particle dynamics (DPD) simulation, which is well known to preserve the hydrodynamics.[45]

Some early studies on phase separation kinetics in BCPs focused on diffusive transport mechanisms and exhibited that the phases of BCP melt evolved into a frozen micro-structure with a series of morphologies.[46,47] However, a few subsequent studies are done later on using DPD[33,34] and molecular dynamics (MD)[48] simulations to assess the effect of hydrodynamics on the microphase separating BCPs. They observed that the hydrodynamics plays a crucial role by accelerating the dynamic growth of separating copolymers at early times to attain the correct frozen morphologies. The segregation in BCPs at initial times are analogous to spinodal decomposition in polymeric fluids.[48] At late times when domain size becomes comparable to polymer chain length, the topological constraints become significant, and morphologies settle to the frozen micro-structures. Thus, a crossover in the growth law from a power-law behavior: $\ell(t) \sim t^{1/3}$ to a constant value balanced with the frozen microphase structure is observed.[48]

In this work, however, we turn our attention to study the phase separation kinetics of BCP blend in $d = 3$ by exploiting the light stimulated bond cleavage to modify the evolved morphologies. BCP is comprised of linear, long-chain molecules ($A_nB_m$) having two sub-chains consisting of $n$ $A$-type monomers followed by $m$ $B$-type monomers and connected by a covalent bond. $A$ and $B$ monomers are regarded as incompatible, enabling the phase separation when quenched below the miscibility gap.[27,29] The covalent bond connecting both the sub-chains is assumed to be photosensitive, and thus breaks in the light-on state;[7,10,13] the system then behaves as a binary (*AB*) polymer melt and undergoes macrophase separation when quenched below the critical temperature.[45,49] Conversely, in the light-off state, phase separation cannot advance to macroscale due to the covalent bond between *A* and *B* immiscible sub-chains. Instead, micro-domains rich in either of the *A* and *B* components are formed.[33,34,48] More details to follow later.

Most of the previous reaction-induced phase separation studies are performed either to find the equilibrium properties of the system or based on the coarse-grained models that use uncontrolled approximations to model the velocity field.[18,19,50–52] To our knowledge, there is no particle-based simulation study reported on the kinetics of microphase separation in BCPs involving photo-controlled reactions during *on* and *off* states of the light. However, since there is no direct use of light, the same model can also be used to break bonds due to the stretching of the simulation box or other deformations in the system.[53–55]

Here we use the DPD framework to simulate the critical BCP melt ($A_nB_n$), which is a mesoscale particle method that bridges the gap between macroscopic and microscopic simulations. The particle-based DPD approach has the advantage of naturally integrating flow fields and preserving the hydrodynamics in the system. Combining with coarse-grain techniques, DPD is very suitable for simulating gaseous or simple/polymeric fluid systems.[33,34,45,56] It has been successfully applied to several areas of interest, mainly to simulate the equilibrium and dynamical properties of polymers melt[33,34], the polymer in solution[45], and polymer gels[13,56,57]. In particular, we adopt the recently developed computational approach based on the DPD framework[56] to simulate the controlled radical polymerization (CRP) within polymer networks in the presence of moieties. The DPD model[56] was validated with the prior experimental and modeling studies.[58] The same framework was adopted further to develop the first simulation approach for modeling photo-CRP within polymer networks.[7,10,13] This DPD model allowed us to systematically examine the relative effects of the photo-initiation (due to bond breaking), propagation with monomers and crosslinker, and termination reactions in the formation of nanocomposite gels[13] and self-healing gels[7,10] that were consistent with prior experimental data[8,9,12], and hence, obtain greater insight into the polymerization process.

Below, we start by describing our computational model and methodology that captures the effect of alternate light on and off cycles on phase separation dynamics of BCP melt. We present the simulation results and discussions in Sec. 3 and finally, Sec. 4 concludes this paper with a summary.

## 2. Model and simulation details

Dissipative particle dynamics (DPD) proved to be a robust approach to simulate the dynamic behavior of diverse, complex systems at mesoscale.[59–64] In the DPD approach, a single particle (or bead) is characterized by a molecule or a collection of particles cooperating through a soft-core potential, and therefore, DPD is well-known to be an advantageous technique to simulate a system over a more substantial length and time scales than a traditional MD simulation. In particular, one can perform DPD simulation of a system with volume up to 100 *nm* in linear dimension and time scale of tens of microseconds.[60,61] Whereas for a classical MD simulation, one may need multi Graphics Processing Unit (GPU) accelerated machines to perform massively parallel scientific computations to simulate a system with the same length and time scales. The particle dynamics is controlled by integrating Newton's equation of motion:[61]

$$\frac{d\boldsymbol{p}_i}{dt} = \boldsymbol{f}_i(t) = \sum_{j \neq i} \left[ \boldsymbol{F}_{ij}^C + \boldsymbol{F}_{ij}^D + \boldsymbol{F}_{ij}^R \right], \qquad (2)$$

where $\boldsymbol{p}_i = m_i \boldsymbol{v}_i$ is the momentum vector of $i^{th}$ bead with mass, $m_i$, and velocity, $\boldsymbol{v}_i = d\boldsymbol{r}_i/dt$. The position vector is denoted by $\boldsymbol{r}_i$, and the mass of each particle is set to $m_i = m$. Three pairwise additive forces, specifically a purely repulsive conservative force ($\boldsymbol{F}_{ij}^C$), dissipative force ($\boldsymbol{F}_{ij}^D$), and random (stochastic) force ($\boldsymbol{F}_{ij}^R$) constitute an effective force $\boldsymbol{f}_i(t)$ acting on each particle $i$. The summation in Eq. (2) runs over all the $j$ beads within a cutoff radius $r_c$ from the $i^{th}$ bead.

The most common choice of the conservative force in DPD is a soft repulsive interaction between the beads[61] $i$ and $j$ which is linear up to the cutoff distance $r_c$, and is given by

$$\boldsymbol{F}_{ij}^C = a_{ij}\left(1 - \frac{r_{ij}}{r_c}\right)\hat{\boldsymbol{r}}_{ij}. \qquad (3)$$

Here, $a_{ij}$ is the maximum repulsion between the beads $i$ and $j$ separated by a distance $r_{ij} = |\boldsymbol{r}_{ij}|$, and $\hat{\boldsymbol{r}}_{ij} = \boldsymbol{r}_{ij}/|\boldsymbol{r}_{ij}|$ determines the force direction. The dissipative and the random force contributions are given by[59,60]

$$\boldsymbol{F}_{ij}^D = -\gamma \omega_D(r_{ij})(\hat{\boldsymbol{r}}_{ij} \cdot \boldsymbol{v}_{ij})\hat{\boldsymbol{r}}_{ij}, \qquad (4)$$

and

$$\boldsymbol{F}_{ij}^R = \sigma \omega_R(r_{ij}) \xi_{ij} \hat{\boldsymbol{r}}_{ij}. \qquad (5)$$

The strength of dissipative (friction coefficient) and stochastic forces are denoted by $\gamma$ and $\sigma$, respectively; $\mathbf{v}_{ij} = \mathbf{v}_i - \mathbf{v}_j$ represents the relative velocity of the beads. Similar to $\mathbf{F}_{ij}^C$, the other two forces also act along $\hat{\mathbf{r}}_{ij}$. To ensure the correct canonical equilibrium state, a system follows the fluctuation-dissipation theorem where the strength of dissipative and random forces are coupled with a relation:[59,60]

$$\sigma^2 = 2\gamma k_B T, \tag{6}$$

and their weight functions are paired with a condition:[61]

$$\omega_D(r_{ij}) = \omega_R(r_{ij})^2 = \left(1 - \frac{r_{ij}}{r_c}\right)^2 \text{ for } r_{ij} < r_c, \tag{7}$$

where $k_B$ denotes the Boltzmann constant, and $T$ is the equilibrium temperature of the system.

The noise amplitude, $\xi_{ij}$ in Eq. (5) symbolizes a Gaussian random variable of zero-mean and unit variance:[59]

$$\langle \xi_{ij}(t) \rangle = 0; \text{ and } \langle \xi_{ij}(t)\xi_{kl}(t') \rangle = \left(\delta_{ik}\delta_{jl} + \delta_{il}\delta_{jk}\right)\delta(t - t'). \tag{8}$$

The symmetry property, $\xi_{ij} = \xi_{ji}$ ensures the local momentum conservation.[59,61] Note that, we choose the most common choice of the functional form of weight function in Eq. (7) which is the same as the form of conservative force in Eq. (3). However, other choices are also permitted as long as the conditions in Eqs. (6-7) are satisfied.[60,63,64] A remarkable advantage of using DPD technique is that each of the pairwise forces in Eqs. (3-5) conserves the momentum locally, and thus, preserves correct hydrodynamic[62] behavior of complex fluids containing only a few hundred particles.[59,61]

*System details and model parameters:* The beads in a BCP chain are connected by harmonic bonds (bead-spring model)[65,66] with potential energy:

$$E_b = 1/2 \, k_b (r - r_0)^2 \tag{9}$$

where $k_b = 128$ is an elastic constant and $r_0 = 0.5$ is an equilibrium bond distance.[10,13,64] The chain stiffness is provided by the angle potential:

$$E_a = 1/2 \, k_a (\cos\theta - \cos\theta_0)^2, \tag{10}$$

where $k_a = 10$ determines the strength of the interaction, $\theta$ is the angle between the successive bonds along a chain itself. The equilibrium value of the angle is set to $\theta_0 = 180$ degree.[66] We set $a_{ij} = 25$ (in the units of $k_B T/r_c$) for the interaction between any two beads of the same component (A − A and B − B interaction in BCP is considered chemically compatible), and $a_{AB} = 60$ for the chemically incompatible beads.[61] These choices of $a_{ij}$ energetically favor phase separation in the system. The time evolution of the system is obtained by integrating the equation of motion (Eq. (2)) using a modified velocity-Verlet algorithm.[61,67] In the simulation, the cutoff radius, $r_c$ introduces a characteristic length scale in the system so that $r_c = 1$ in dimensionless DPD units; $k_B T$ represents the typical energy scale. We set $\gamma = 4.5$ to provide a relatively rapid equilibration to the system temperature and numerical stability for the specified time steps.[10,13] Further, we quench BCP melt at a reduced temperature $T^* = T/T_0 = 1$, where $T_0$ is taken as a reference temperature. Subsequently, we show that this quench temperature is well below the corresponding critical temperature for phase separation to take place. The system is integrated at the time step $\Delta t = 0.02\tau$ where $\tau = (mr_c^2/k_B T_0)^{1/2}$ is defined as the characteristic time scale. The total bead number density is set to $\rho = 3$, which is a reasonable choice for DPD simulation of liquids.[59,61] The dimensional values of length $r_c$ and time $\tau$ are observed to be $0.97\ nm$ and $8.3\ ps$, respectively.[65,68]

Further, to incorporate the effect of light in the framework of DPD simulation, we consider the bond joining two BCP blocks as highly light-sensitive. So the essential reactions that we discuss here are: (i) the *breaking of light-sensitive bond* (initiation reaction in the on-state of light), and (ii) the *reformation of broken bond* (termination by combination reaction when light is off).[7,10,13] We neglect any other reactions such as degenerative chain transfer and coupling of the same radicals (which can prevent the reformation of the covalent bond between *A* and *B* blocks) that may take place in the system.[7,10,13] In the bond-breaking reaction time step, we check all possible bonds joining *A*-type (blue beads) and *B*-type (yellow beads) blocks of BCP chains and assess the bond stretching. The bond-breaking reaction is performed with probability, $P_{bb}$ when the bond length is greater than a cutoff size $r_b = 1.4\ r_0$, where $r_0 = 0.5$ is the average bond length;[7,10,13] We set the cutoff size to be $r_b = 1.4\ r_0$ since any value less than this resulted in significant deviations from first-order kinetics, and the higher values ($r_b \geq 1.6\ r_0$) slow down the bond-breaking reactions. Note that within our DPD model, the change in bond breaking probability ($P_{bb}$) changes the number of broken bonds per time (we will discuss this shortly). This means that variation in $P_{bb}$ mimics the effect of change in light intensity, i.e., at a higher light intensity, more bonds will break and vice-versa.[7,10,13] The bond cleavage transforms the blue and yellow beads, which form the light-sensitive covalent bond, into two active radicals shown by the glowing cyan and orange colors, respectively. In the absence of light, these active radicals contribute to the termination by combination reaction with probability, $P_{bc}$ to regenerate the BCP chain.[10,13] Since these radicals belong to the incompatible blocks, the energetic drive for binding the radicals to regenerate the BCP chain overwhelms the enthalpic difference. Hence, the cyan radical recombine with the orange radical to reform the BCP chain. Note that the living free-radical polymerization techniques are used to form BCPs from incompatible moieties.[69,70]

In each termination step, first, an active radical is selected randomly, and then we check for another active bead-type randomly within a cutoff radius, $r_i$, of the first selected radical. Thus all the band pairs are chosen randomly (and with equal probability) among all pairs in the given cutoff, $r_i$. Following the previous reports on FRP, we set $r_i = 0.7$ to reproduce the correct first-order kinetics of the polymerization process.[13,56] In general, the reacting pair of active beads form a bond with a combination probability, $0 < P_{bc} < 1$. Namely, for each reaction time step, a random number $q \in (0,1)$ is generated and then compared to $P_{bc}$. The reaction is accepted if $q < P_{bc}$ and rejected otherwise. Thus, each successful reaction results in an irreversible covalent bond formation. The time step for each reaction is set to $\tau_r = 0.2\tau$, i.e., each reaction is performed every 10 DPD time step.[10,54,56] At the onset of DPD simulation, the box size is set to $L_x \times L_y \times L_z = 64 \times 64 \times 64$ (dimensionless units). The periodic boundary conditions are applied in all three directions. Since number density in the system is set to $\rho = 3$, total DPD beads in the system are $N = \rho \times L_x \times L_y \times L_z = 786432$. A high number density ensures that the system remains far away from the gas-liquid transition during the simulation. We carry out a comprehensive study on utilizing external stimuli such as light to alter the microphase separation kinetics of BCP melt[48] for a critical composition ($N_A : N_B = 1:1$) where $N_A$, and $N_B$ indicate the number of $A$-type, and $B$-type beads such that $N = N_A + N_B$. The length of a BCP chain is considered to be $L_p = 32$ with each block contains $N_A = N_B = 16$ beads (see Fig. 1c). Thus, the total number of BCP chains of length ($L_p$) in the system is set to $N_{pc} = N/L_p = 24576$. To study the system under consideration, we use DPD module of LAMMPS simulation package.[67,71]

To understand the effect of light, we perform two different sets of experiments. In set 1, phase separation begins with the light off (cycle 1) and then followed by alternating on and off cycles of the equal period ($t = 600$) up to cycle 5 (off → on → off → on → off); here, $t =$ number of simulation steps × $\Delta t$. However, in set 2, we choose opposite cycles of set 1 with the same period (on → off → on → off → on). Here, we vary $P_{bb}$ to highlight the influence of light intensity on the structure of evolving morphologies, growth law, and scaling functions. Mainly, we perform DPD simulations for three different values of $P_{bb} = 0.1, 0.5,$ and $1.0$ by keeping bond combination probability, $P_{bc} = 1.0$.

*Morphology characterization function and length scale:* Amid many, the two most essential characterization functions for the evolution morphology are the two-point ($\boldsymbol{r} = \boldsymbol{r_1} - \boldsymbol{r_2}$) equal-time correlation function, $C(r,t)$, and its Fourier transform, the structure factor, $S(k,t)$.[27,29] The correlation function is defined as

$$C(r,t) = \langle \psi(r_1,t)\psi(r_2,t)\rangle - \langle \psi(r_1,t)\rangle\langle \psi(r_2,t)\rangle, \tag{11}$$

where $\psi(r_1,t)$ is the order parameter; it is defined as the local concentration difference of the distinctive constituents on a discrete lattice site $r_1$ at a given $t$. The angular brackets denote an ensemble average obtained over five independent runs. The structure factor is defined as

$$S(k,t) = \int dr\, e^{ik\cdot r}\, C(r,t), \tag{12}$$

where $k$ represents the scattering wave-vector.

We divide the simulation box into non-overlapping boxes of unit size and count the number of $A$-type beads ($n_A(r,t)$), and $B$-type beads ($n_B(r,t)$) in each of the boxes. The coarse-grained order parameter at a lattice point $r$ is computed as:[28,45,49]

$$\psi(r,t) = \frac{n_A(r,t) - n_B(r,t)}{n_A(r,t) + n_B(r,t)}. \tag{13}$$

Thus, the continuum bead configurations are mapped onto a discrete simple cubic lattice of size $64 \times 64 \times 64$. When $n_A(r,t) = n_B(r,t)$ in a unit box, the corresponding order parameter is assigned a value as $0 < \psi(r,t) < 1$ or $-1 < \psi(r,t) < 0$ with equal probability.[28] To see the Porod-tail[72] in $S(k,t)$ at larger $|k|$ values, we "harden" the order-parameter field such that $\psi = +1$ for $\psi > 0$, and $\psi = -1$ for $\psi < 0$.[45,49] Note that only higher-order Bragg-reflections are effectively visible in the domain structure, i.e., at $(2n+1)k_m$ for $n = 1, 2, \cdots$, where scattering around the magnitude of the most probable wave vector, $k_m$ dominates the structure factor spectrum that leads to a narrow peak. A distinct set of wave vectors that remain physically significant for a finite lattice is given by

$$k = \left(\frac{2\pi n}{L}\right) \text{ with } n = (n_x, n_y, n_z), \tag{14}$$

where $0 \leq n_i \leq L$ for $i = x, y, z$.[45,73] The ordering in segregated phases is identified by the *number of peaks* in the structure factor obtained from the periodicity of domains for all the $k's$. Since we consider an isotropic system, the statistics of both $C(r,t)$ and $S(k,t)$ can be improved by spherical averaging; corresponding spherically averaged quantities are denoted by $C(r,t)$ and $S(k,t)$, respectively.[45,73]

An established feature of phase separation kinetics is the presence of a characteristic length scale (average domain size), $R(t)$, defined uniquely for the isotropic system. The dynamical scaling in the correlation function and the structure factor due to the unique length scale in the system has the following forms:[27,29]

$$C(r,t) = f(r/R(t)), \quad (15)$$

$$S(k,t) = R(t)^d g(kR(t)), \quad (16)$$

where $f(r/R(t))$ and $g(kR(t))$ are the scaling functions. The average domain size, $R(t)$, can be estimated exploiting the scaling properties of $C(\boldsymbol{r},t)$ and $S(\boldsymbol{k},t)$ as specified in Eqs. (11-12). There are a few appropriate definitions for computing $R(t)$ such as (i) the distance at which the correlation function decays to some fraction of its maximum value ($C(0,t) = 1$), (ii) the inverse of the first moment of $S(k,t)$, and (iii) first moment of the normalized domain-size distribution function. Since all these definitions are equivalent in the scaling regimes (*i.e.*, they differ only by constant multiplicative factors), we observed that the decay of $C(r,t) \to 0.2$ gives a good measure of $R(t)$.[74,75] Typically, the average domain size follows the power-law growth: $R(t) = R_0 + at^\phi$ where $R_0$ is considered to be a length scale of transient growth regime in the system since the quench[45,76] and $a$ is the constant fitting parameter. Therefore, to extract the growth exponent $\phi$, we plot $R(t)$ versus time, $t$ on a logarithmic scale.

### 3. Results and discussions

It is a well-established fact that the domain coarsening of various phases of a multicomponent fluid when quenched below the critical temperature at time $t = 0$, is a scaling phenomenon where morphologies are self-similar at two different times. In this section, we present the details of the numerical results of phase separation kinetics in BCP melt for both the sets at various cycles, as mentioned above. The initial homogeneous configuration for all the cases discussed here are prepared by equilibrating the system for $t = 2000$ at a high-temperature $T = 10$ (see Fig. 1b). Next, we quench the system to a lower temperature, $T = 1$, and reset the time measure at $t = 0$ and observe the domain evolution at various times.

At the outset, we first explore the earlier studies on segregating critical binary ($AB$) polymer melt and critical BCP melt in $d = 3$ to certify the parameter values used in the simulation.[45,48,49] The simulation box size and number density of beads are the same as discussed earlier (*i.e.*, $N = 3 \times 64^3$; polymer chain length $L_p = 32$). The domain evolution of $A$-rich and $B$-rich phases of the polymer melt is depicted in Fig. S1 at times, $t = 500, 1000, 1500,$ and $2000$. The time dependence of the average domain size shows a crossover from the viscous hydrodynamic ($\phi \sim 1$ till $t < 100$) to the inertial hydrodynamic ($\phi \sim 2/3$) regimes. The solid and dashed lines display the expected growth exponents in various growth regimes. Since DPD is best known to preserve the hydrodynamic behavior in the system, we observe that the diffusive growth ($\phi \sim 1/3$) regime remain short-lived on the time scale of our simulation.[59,61] The excellent data collapse of the scaling functions (inset of Fig.

S1) indicate that the evolving systems belong to the same dynamical universality class.

Next, the phase-separated domains of BCP melt are illustrated in Fig. S2 at $t = 1000, 2000, 3000,$ and $5000$. Due to microphase separation in BCP melt, we correctly noticed the expected diffusive growth ($\phi \sim 1/3$) regimes at early times (*i.e.,* $t < 200$) that further saturates to an average periodic domain size in the system. The data overlap of the characteristic functions at various times confirm the presence of dynamical scaling (see the inset). The data oscillation around $C(r,t) = 0$, and a secondary peak (bump) in $S(k,t)$ at larger $kR(t)$ other than the main peak, confirms the formation of periodic domain structures in microphase separated BCP melt when quenched below the critical temperature. Thus, the above description verifies the correct modeling of phase separation dynamics for both the polymer melt and BCP melt systems.

Henceforth, we focus on to phase separation dynamics in BCP melt ($A_n B_n; n = 16$) by employing alternate switching of light between "on" and "off" states that lead to bond breaking and bond reformation reactions. At first, we study the case where switching of light in various states take place as off → on → off → on → off (we call this set 1). We consider the first cycle for $t = 600$ to make sure that the system is reached in the scaling regimes before we start the other cycles. (From Figs. S1 and S2, an inertial hydrodynamic growth regime is noted for polymer melt, and the pinning of growth for BCP melt before $t = 600$.) A standard microphase separation occurs when the first cycle remains in an off state, and thus no bond reformation reaction takes place either due to the absence of active radical in this cycle. The domain morphologies of BCP melt after completing each period for set 1 is exhibited in Fig. 2; the bond-breaking probability is fixed to $P_{bb} = 1.0$ for the on-cycles, and for the off-cycles, bond-formation probability (termination of active radicals by combination) is set to $P_{bc} = 1.0$. The blue and yellow beads represent *A*-rich and *B*-rich domains, respectively, in Fig. 2. The bond-breaking reaction starts in cycle 2 (on state) of set 1. Note that depending on the values of $P_{bb}$, the kinetics gradually moves from microphase separation in BCP melt ($P_{bb} = 0$) to macrophase separation in binary polymer melt ($P_{bb} = 1$); this is due to the removal of bond constraint connecting both unfavorable *A* and *B*-blocks in BCP melt (see Fig. S3). The visuals, in Fig. S3, illustrate the larger domain size for $P_{bb} = 1.0$ than $P_{bb} = 0.1$ where the bond-breaking is slower than earlier.

When phase separation begins with the light on (*e.g.,* in cycle 1 for the set 2), bond breaking gradually turns BCP melt into a binary polymer melt, and thus phase separation takes place via spinodal decomposition[27,29] as in Fig. S4. The values of $P_{bb}$ determine the rate of domain growth. Again, through off-cycle 2 of set 2, active radicals created due to the bond cleavage in cycle 1 undergo termination reaction by combination ($P_{bc} = 1.0$) to regenerate the broken covalent bonds (see the morphologies in Fig. S4 at $t = 1200$). We observe the different phase-separated domain patterns during cycle 2 at the same, $P_{bc} = 1$ because of the distinct domain patterns formed during cycle 1 at different $P_{bb}$. Later on, we show that typically, the average size of the domain, $R(t)$, is more significant when phase separation initiates with the light on with $P_{bb} = 1.0$.

The total number of photosensitive covalent bonds joining both the unfavorable blocks in BCP melt is $N_{pc} = 24576$. To demonstrate the effect of light, we plot the normalized cumulative number of bonds cleaved ($N_{bb}$) during on-cycles and recombined ($N_{bc}$) during off-cycles for both the sets in Fig. 3. The black and green curves in Fig. 3a exhibit the number of bonds broken during the on-cycles of set 1 at $P_{bb} = 1.0$. Notice that $N_{bb} = N_{pc}$ in cycle 2 (black curve) whereas in cycle 4, $N_{bb} \lessapprox N_{pc}$ (green curve), a reason could be the trapping of active radicals within the domain matrix during cycle 2 causes the lesser number of covalent bond reformation, $N_{bc} < N_{pc}$ (red curve) in cycle 3, and thus a lower number of bonds breaking during cycle 4. An increase in the probability value, $P_{bb}$ enhance the rate of bond-breaking during on sequences as represented by the black ($P_{bb} = 0.1$), red ($P_{bb} = 0.5$), and green ($P_{bb} = 1.0$) curves, respectively in Fig. 3b during cycle 2 for set 1. Thus, our model precisely captures the effect of variation in $P_{bb}$. A similar observation is made for all the sequences in set 2, as plotted in Fig. 3c.

The dashed lines in each plot demonstrate the exponential fit, $N_{bb}(t)/N_{pc} \sim (1 - e^{-\lambda P_{bb} t})$, where $\lambda$ is a fitting parameter for the corresponding simulated data shown by various symbols in the same color, and $t$ = number of simulation steps × $\Delta t$ (= $0.02\tau$). The value of $\lambda$ for the dashed lines in fig. 3b are $0.1, \ 0.092$, and $0.097$, for $P_{bb} = 0.1, \ 0.5$, and $1.0$, respectively. We observed that the fitting parameter could be defined as $\lambda = \Delta t/\tau_r \simeq 0.1$ which remain constant for different $P_{bb}$ values. So, the variaton in $P_{bb}$ and $\tau_r$ parameter values can efectively control the rate of bond-breaking reactions. We define the bond-breaking rate constant as $k = \lambda P_{bb} = 0.01, \ 0.05,$ and $0.1$, for $P_{bb} = 0.1, \ 0.5,$ and $1.0$, respectively.[54] Therefore, in our scheme, the bond-breaking reaction occurs every $k^{-1} = 100, 20,$ and $10$ (units of $\tau$) time intervals for the given $P_{bb}$ values. Thus, the bond-breaking reactions are slower than the characteristic diffusion time scale of DPD simulation and hence belongs to a kinetically controlled growth regime.[54] These data fit to show that our model reproduces the first-order kinetics of bond-breaking reactions.[10,13,54] Since there is no direct effect of light intensity in the framework of our simulation, however, the variation in bond breaking rate constant, $k$ mimics an impression of variation in light intensity.

In Fig. 4a, we plot the radial distribution function (RDF), $g_{AB}(r)$, to characterize the morphologies displayed in Fig. 2. Notably, we compute the local density ratio of $A$-type (blue) bead being at a distance, $\boldsymbol{r}_{ij} = (x_{ij}, y_{ij}, z_{ij})$ around a $B$-type (yellow) bead. Note that the upper bound of $x_{ij}, y_{ij}$, and $z_{ij}$ is less than or equal to half of the periodic box length.[77] The oscillation in the black curve (multiple peaks) verifies the formation of typical periodic domains in BCP melt during off-cycle 1. The green and purple curves do not show any second peak in $g_{AB}(r)$ as one may expect during the remaining off-cycles too. Since we are well aware of the fact that microphase separation in BCP is a slow dynamic process that requires a longer time to develop the periodic domains. Hence, one possible reason for not showing a second peak in $g_{AB}(r)$ during the remaining off-cycles could be the assigned shorter time for domain evolution once the bonds are reformed. The peak position in $g_{AB}(r)$ shifts to a more considerable value of $r$ as domains grow further with time. To verify the domain coarsening displayed in Fig. 2, we plot the spatial number density distributions of $A$-beads along the transverse ($z$) direction in Fig. 4b. The plots indicate the increase in height and width of the number density curve with time and confirm the growth of $A$-rich domains. The symbols and their colors in Fig. 4b represent the same time steps as in Fig. 4a.

In the light on cycles (cycle 2 & 4), covalent bonds cleaved at a faster rate for the higher value of bond breaking probability (high light intensity). Hence, the system rapidly evolves as a binary ($AB$) polymeric fluid mixture having chain length, $L_{pc} = 16$. Therefore, the shift in the peak position of $g_{AB}(r)$ to a higher $r$ at $P_{bb} = 1.0$ confirms the presence of a larger average domain size (see the green curve in Fig. 5). The black and red curves show RDFs at $P_{bb} = 0.1$ and $P_{bb} = 0.5$, respectively, with the peak positions at corresponding lower $r$ values. An analogous behavior of RDF curves is observed at different $P_{bb}$ values during each on-cycle for the set 2. Nevertheless, the bond combination probability is set to $P_{bc} = 1.0$ during each off-cycles; $g_{AB}(r)$ curves show an exact overlap of the peak positions, *i.e.*, the same average domain size (the plot is not displayed here). Any difference at all observed in domain sizes during the previous on cycle at different $P_{bb}$ values become insignificant in the given time limit of the off-cycles.

We plot the scaling functions, the correlation function, $C(r,t)$ vs. $r/R(t)$, in Fig. 6a and the structure factor, $S(k,t)$ vs. $kR(t)$, in Fig. 6b. In these plots, we examine how the evolution morphologies depicted in Fig. 2 are governed by the alternate on and off cycles; this comparison of the scaling functions are made at the end of each period (i.e., at every $t = 600$; denoted by specified symbol types) when the system is already in the scaling regimes. A significant deviation from the scaling is noted between both the scaling functions at $t = 600$ (black curve; off-cycle 1) and $t = 1200$ (red curve; on-cycle 2) in Fig. 6. This is quite apparent as the kinetics through both the cycles are very different. The off-cycle 1 generates the usual microphase-structures of BCP melt (no reaction involved) characterized by a more oscillatory $C(r,t)$ (black curve in Fig. 6a) and a bump (second peak) at a sizeable $k$ value other than the principal peak in $S(k,t)$ at small $k$ (black curve in Fig. 6b); this is a characteristic of a periodic morphology.[46] Whereas on-cycle 2 causes a macrophase-structures due to spinodal decomposition in binary polymer melt with a chemically-controlled regime (bond-breaking reactions) for a few initial time steps. Interestingly, a significant data collapse of both the scaling functions in cycle 3 onwards (the green, blue and purple curves) suggest that the morphologies are equivalent and their statistical properties are independent of on- and off-state of the light at late times and thus belong to the same universality class. One possible reason for this could be the formation of large domains during the on-cycles. Therefore, the assigned duration for t off-cycles, when a relatively much slower microphase separation kinetics in BCP melt (due to quick restoration of broken bonds in off-cycles) takes place, is not sufficient enough to affect the length scale of the system and hence the scaling functions. However, $S(k,t)$ shows the expected power-law decay ($\sim k^{-4}$) of the tail in Fig. 6b; this is known as Porod's law: $S(k,t) \sim k^{-(d+1)}$ for $k \to \infty$ as a result from the scattering off sharp interfaces, where $d = 3$ is the system dimensionality.[72,78]

Further, the nature of dynamic scaling functions within the cycle is illustrated in Fig. S5 for the set 1, and in Fig. S6 for the set 2. Specifically, we focus on the first two cycles of both the sets in which the light stimulated chemical reaction is involved keeping $P_{bb} = 1.0$ and $P_{bc} = 1.0$, *i.e.,* cycles 2 and 3 for the set 1, and cycles 1 and 2 for the set 2. These two figures clearly indicate that the first on-cycle has a significant impact on the scaling functions. Notice that the scaling functions illustrate the crossover at early times; thus, the deviation from scaling is seen when bond-breaking reactions control the kinetics as shown by the black and red curves at the corresponding times in Figs. S5ab and S6ab. Though, at late-times, the scaling is fully preserved when the kinetics is analogous to spinodal decomposition in the polymer melt, shown by the green and blue curves.

Nevertheless, the scaling functions show an excellent data collapse during the first light off state when it comes after the first on-state, e.g., off-cycle 3 for set 1 (see Figs. 5Scd), and off-cycle 2 for set 2 (see Figs. 6Scd). This observation suggests that the evolving morphologies after the first on-cycle are part of the same dynamical universality class irrespective of the "on" or "off" state of the light.

In Fig. 6c, we compare the scaling function $C(r,t)$ versus $r/R(t)$ and in Fig. 6d, we plot the corresponding structure factor, $S(k,t)R(t)^{-3}$ versus $kR(t)$ at the end of each cycle for set 2 (on → off → on → off → on). A deviation from the scaling is observed for data at $t = 600$ (on cycle; black symbol). However, the well-fitted scaling functions, illustrated by the red ($t = 1200$), green ($t = 1800$), blue ($t = 2400$), and purple ($t = 3000$) curves, are observed at the end of each cycle for set 2 after evolving the system in the light on state. This further confirms our above observations that the domain growth beyond the first on-cycle is part of the same dynamical universality class irrespective of the "on" or "off" state of the light. The solid line of slope $-4$ indicates the well-known Porod-tail behavior of the structure factor.

Finally, we report the domain growth law. We plot $R(t)$ *vs.* $t$ on a logarithmic scale in Fig. 7 for the evolution through each cycle for the set 1. The black curve in Fig. 7a shows the apparent microscopic domain growth of $A$ and $B$-blocks for BCP blend during cycle 1 (off). A solid line represents the expected diffusive growth exponent ($\phi \sim 1/3$) for a limited time window. The black curve then crosses over to the saturated microphase morphology. When bonds between the $A$ and $B$-blocks of BCP chain breaks during the second cycle (on), the system behaves as a binary ($AB$) polymeric blend, and hence, a macroscopic domain growth is noted. The black curve in Fig. 7b displays a gradual crossover to the inertial hydrodynamic growth ($\phi \to 2/3$) regime. This agrees with the growth regimes anticipated at a late-stage for the polymeric fluid mixture, as presented in Fig. S1. We observe that the viscous hydrodynamic growth ($\phi \sim 1$) regime is very short-lived in this case is due to the already attained length scale in cycle 1. The domain growth observed during off cycles 3 (red curve) and 5 (green curve) in Fig. 7a follows the similar trend of slow growth as in off-cycle 1; however, the saturation in length scale has not been achieved in the given time window due to already attained macroscopic domain in the previous on-cycle. Similarly, the red curve in Fig. 7b displays a relatively slower domain growth during on-cycle 4 due to already achieved inertial hydrodynamic growth in the previous on cycle.

The phase separation for set 2, begins with on-cycle; hence we note both the expected hydrodynamic growth regimes, $\phi \sim 1$ and $\phi \sim 2/3$, as shown by the black curve in Fig. 8a. Whereas during cycle 2 (off), we notice the usual crossover to diffusive growth regimes ($\phi \to 1/3$) at late times, as depicted by the black curve in Fig. 8b. However, in other cycles, the crossover to expected growth regimes: inertial hydrodynamic growth ($\phi \to 2/3$) during on cycles 3 and 5 (red and green curves in Fig. 8a), and diffusive growth ($\phi \to 1/3$) during off-cycle 4 (red curve in Fig. 8b), are slower due to already attained macroscopic growth in the first on cycle.

## 4. Conclusions

We utilized the dissipative particle dynamics (DPD) technique to simulate the evolution of phase-separating critical block copolymer (BCP) melt by passing it through alternate "on" (bond breaking) and "off" (bond combination) cycles. We discussed its effect on evolution morphologies, scaling functions, and length scales for the two different sets.

We found that the rate of bond-breaking during the on-state sequences enhanced with the increase in $P_{bb}$. For a given value of the rate constant, $k$, the number of broken bonds were consistent with the number of light-sensitive bonds in the system during the first cycle. Our model correctly encapsulates the effect of change in the rate constant, $k$ that mimics an impression of variation in light intensity. In other on-state cycles, however, the number of broken bonds slightly reduced due to the lesser covalent bond formation via reversible radical deactivation in the previous off cycles, which happened possibly because of the trapping of active radicals within the domain matrix.

The shift in the peak positions of $g_{AB}(r)$ to a higher $r$ further demonstrates that a higher $k$ (that mimics higher light intensity) leads to a more segregated domain; this occurs as macrophase separation dynamics begin early due to quick bond breaking at higher $k$. Nonetheless, the probability of bond combination is set to 1.0 during all the off cycles. $g_{AB}(r)$ curves show an exact overlap of the peak positions for $P_{bb}$ values in the previous on-cycles. Interestingly, any difference in the peak position observed in the previous on cycle at three different $P_{bb}$ values nullified in the given time limit of the off-cycles.

For set 1, the scaling functions $C(r,t)$ and $S(k,t)$ obtained at the end of the first two cycles exhibited a significant deviation from the master function as the kinetics in both the sequences were very different. Remarkably, in cycle 3 onwards, we observed an excellent data collapse of both the scaling functions; this suggests that the morphologies are statistically similar and independent of on- and off-state of the light at late times, and thus belong to the same universality class. For set 2, where the evolution started with an on cycle, well-fitted scaling functions were noted at the end of each cycle. Overall, we found that the first on-cycle determines the behavior of scaling functions for the rest of the sequences. The domain growth after the first on-cycle integrated to the same universality class irrespective of the state of light for the given period. Similarly, the scaling functions within a cycle illustrate the deviation at early times until the first on-cycle when bond-breaking reactions control the kinetics.

We further observed that the average domain size, $R(t)$ is larger when phase separation instigated with the light on-state; the domain size becomes even more prominent at a higher light intensity. Furthermore, we accessed the crossover from viscous to the inertial hydrodynamic regime ($R(t) \sim t^1 \rightarrow t^{2/3}$) when quenched with the on-cycle. Since the length scale has already crossed over to the inertial hydrodynamic regime, the following on-cycles also illustrated the length scale gradually approaching the same growth regimes, $\phi \rightarrow 2/3$. Nevertheless, when the system is quenched with off-cycle, following on-cycles also exhibited the inertial hydrodynamic regimes only at late times with a broader transient regime at early times of the corresponding on-cycle. Whereas, during off-cycles, the length scale remained consistent with the gradual crossover to diffusive growth ($\phi \rightarrow 1/3$) regime in a limited time window.

In the future work, we plan to investigate the light-controlled segregation kinetics in BCP melt and BCP solutions with symmetric and asymmetric block compositions. Explicitly, we intend to study the effect of keeping a sufficiently long off-cycles on evolution morphologies between the short period of on-cycles. Given the enormous technological importance of the morphologies obtained in various BCPs, we firmly believe that these results will incite additional attention and offer a broad framework to analyze the simulation and experimental works on the stimuli-responsive growth of nanostructures in BCP melt.

## Acknowledgments

A.K.S is grateful to CSIR, New Delhi, India, and A.C. thanks to IIT (BHU) for financial support. A.S. acknowledges the financial support from SERB Grant No.: ECR/2017/002529 by the Department of Science and Technology, New Delhi, India.

## Conflicts of interest

There are no conflicts of interest to declare.

## References

1  Y. Luo and M. S. Shoichet, *Nat. Mater.*, 2004, **3**, 249–254.

2  A. Lendlein, H. Jiang, O. Jünger and R. Langer, *Nature*, 2005, **434**, 879–882.

3  T. F. Scott, A. D. Schneider, W. D. Cook and C. N. Bowman, *Science*, 2005, **308**, 1615–1617.

4  C. J. Kloxin, T. F. Scott, H. Y. Park and C. N. Bowman, *Adv. Mater.*, 2011, **23**, 1977–1981.

5    Y. Amamoto, J. Kamada, H. Otsuka, A. Takahara and K. Matyjaszewski, *Angew. Chemie*, 2011, **123**, 1698–1701.

6    Y. Amamoto, H. Otsuka, A. Takahara and K. Matyjaszewski, *ACS Macro Lett.*, 2012, **1**, 478–481.

7    M. Chen, Y. Gu, A. Singh, M. Zhong, A. M. Jordan, S. Biswas, L. T. J. Korley, A. C. Balazs and J. A. Johnson, *ACS Cent. Sci.*, 2017, **3**, 124–134.

8    H. X. Zhou and J. A. Johnson, *Angew. Chemie-International Ed.*, 2013, **52**, 2235–2238.

9    M. Chen and J. A. Johnson, *Chem. Commun.*, 2015, **51**, 6742–6745.

10    A. Singh, O. Kuksenok, J. A. Johnson and A. C. Balazs, *Soft Matter*, 2017, **13**, 1978–1987.

11    D. Konkolewicz, K. Schröder, J. Buback, S. Bernhard and K. Matyjaszewski, *ACS Macro Lett.*, 2012, **1**, 1219–1223.

12    M. Chen, M. Zhong and J. A. Johnson, *Chem. Rev.*, 2016, **116**, 10167–10211.

13    A. Singh, O. Kuksenok, J. A. Johnson and A. C. Balazs, *Polym. Chem.*, 2016, **7**, 2955–2964.

14    M. Chen, M. J. MacLeod and J. A. Johnson, *ACS Macro Lett.*, 2015, **4**, 566–569.

15    M. Seul and D. Andelman, *Science*, 1995, **267**, 476–483.

16    Q. Tran-Cong-Miyata and A. Harada, *Phys. Rev. Lett.*, 1996, **76**, 1162–1165.

17    Q. Tran-Cong-Miyata and H. Nakanishi, *Polym. Int.*, 2017, **66**, 213–222.

18    S. Puri and H. L. Frisch, *J. Phys. A. Math. Gen.*, 1994, **27**, 6027.

19    S. C. Glotzer, E. A. Di Marzio and M. Muthukumar, *Phys. Rev. Lett.*, 1995, **74**, 2034–2037.

20    S. Puri and H. L. Frisch, *Int. J. Mod. Phys. B*, 1998, **12**, 1623–1641.

21    Y. Matsushita, *Macromolecules*, 2007, 40, 771–776.

22    Q. Tran-Cong-Miyata, S. Nishigami, T. Ito, S. Komatsu and T. Norisuye, *Nat. Mater.*, 2004, **3**, 448–51.

23    T. Hashimoto, M. Hayashi and H. Jinnai, *J. Chem. Phys.*, 2000, **112**, 6886–6896.

24    M. Hayashi, H. Jinnai and T. Hashimoto, *J. Chem. Phys.*, 2000, **112**, 6897–6909.

25    J. E. G. Lipson, *Macromolecules*, 1991, **24**, 1334–1340.

26    A. Onuki, *Phase Transition Dynamics*, Cambridge University Press, Cambridge, Massachusetts, 2002.

27    A. J. Bray, *Adv. Phys.*, 1994, **43**, 357–459.

28	A. Singh and S. Puri, *Soft Matter*, 2015, **11**, 2213–2219.

29	S. Puri and V. K. Wadhawan, *Kinetics of phase transitions*, CRC Press, Boca Raton, FL, 2009.

30	M. C. Orilall and U. Wiesner, *Chem. Soc. Rev.*, 2011, **40**, 520–535.

31	S. Förster and T. Plantenberg, *Angew. Chemie Int. Ed.*, 2002, **41**, 688–714.

32	J. K. Kim, S. Y. Yang, Y. Lee and Y. Kim, *Prog. Polym. Sci.*, 2010, 35, 1325–1349.

33	R. D. Groot and T. J. Madden, *J. Chem. Phys.*, 1998, **108**, 8713–8724.

34	R. D. Groot, T. J. Madden and D. J. Tildesley, *J. Chem. Phys.*, 1999, **110**, 9739–9749.

35	F. S. Bates, *Science*, 1991, **251**, 898–905.

36	M. W. Matsen and M. Schick, *Phys. Rev. Lett.*, 1994, **72**, 2660–2663.

37	C. C. Honeker and E. L. Thomas, *Chem. Mater.*, 1996, 8, 1702–1714.

38	J. DeRouchey, T. Thurn-Albrecht, T. P. Russell and R. Kolb, *Macromolecules*, 2004, **37**, 2538–2543.

39	I. M. Lifshitz and V. V. Slyozov, *J. Phys. Chem. Solids*, 1961, **19**, 35–50.

40	E. D. Siggia, *Phys. Rev. A*, 1979, **20**, 595–605.

41	T. Hashimoto, M. Itakura and H. Hasegawa, *J. Chem. Phys.*, 1986, **85**, 6118–6128.

42	Y. C. Chou and W. I. Goldburg, *Phys. Rev. A*, 1979, **20**, 2105–2113.

43	C. M. Knobler and N.-C. Wong, *Phys. Rev. A*, 1981, **24**, 3205–3211.

44	F. S. Bates and P. Wiltzius, *J. Chem. Phys.*, 1989, **91**, 3258–3274.

45	A. Singh, A. Chakraborti and A. Singh, *Soft Matter*, 2018, **14**, 4317–4326.

46	Y. Oono and Y. Shiwa, *Mod. Phys. Lett. B*, 1987, **01**, 49–55.

47	Y. Oono and M. Bahiana, *Phys. Rev. Lett.*, 1988, **61**, 1109–1111.

48	A. Singh, R. Krishnan and S. Puri, *Europhys. Lett.*, 2015, **109**, 26006.

49	A. Singh, S. Puri and C. Dasgupta, *J. Chem. Phys.*, 2014, **140**, 244906.

50	P. Dayal, O. Kuksenok and A. C. Balazs, *Langmuir*, 2008, **24**, 1621–1624.

51	R. D. Travasso, O. Kuksenok and A. C. Balazs, *Langmuir*, 2006, **22**, 2620–2628.

52	O. Kuksenok, R. D. M. Travasso and A. C. Balazs, *Phys. Rev. E*, 2006, **74**, 011502.

53	J. Lei, S. Xu, Z. Li and Z. Liu, *Front. Chem.*, 2020, **8**, 115.

54	V. Palkar, C. K. Choudhury and O. Kuksenok, *MRS Adv.*, 2020, **5**, 927–934.

55	E. Bering and A. S. De Wijn, *Soft Matter*, 2020, **16**, 2736–2752.

56   X. Yong, O. Kuksenok, K. Matyjaszewski and A. C. Balazs, *Nano Lett.*, 2013, **13**, 6269–6274.

57   S. Biswas, A. Singh, A. Beziau, T. Kowalewski, K. Matyjaszewski and A. C. Balazs, *Polymer*, 2017, **111**, 214–221.

58   H. F. Gao, P. Polanowski and K. Matyjaszewski, *Macromolecules*, 2009, **42**, 5925–5932.

59   P. Espanol and P. Warren, *Europhys. Lett.*, 1995, **30**, 191–196.

60   P. Espanol and P. B. Warren, *J. Chem. Phys.*, 2017, 146, 150901.

61   R. D. Groot and P. B. Warren, *J. Chem. Phys.*, 1997, **107**, 4423.

62   R. D. Groot, *J. Chem. Theory Comput.*, 2006, **2**, 568–574.

63   P. Nikunen, M. Karttunen and I. Vattulainen, *Comput. Phys. Commun.*, 2003, **153**, 407–423.

64   P. Nikunen, I. Vattulainen and M. Karttunen, *Phys. Rev. E*, 2007, **75**, 036713.

65   C. Junghans, M. Praprotnik and K. Kremer, *Soft Matter*, 2008, **4**, 156–161.

66   H.-P. Hsu and K. Kremer, *J. Chem. Phys.*, 2016, **144**, 154907.

67   S. Plimpton, *J. Comput. Phys.*, 1995, **117**, 1–19.

68   V. Symeonidis, G. E. Karniadakis and B. Caswell, *J. Chem. Phys.*, 2006, **125**, 184902.

69   D. A. Shipp, J. L. Wang and K. Matyjaszewski, *Macromolecules*, 1998, **31**, 8005–8008.

70   K. Matyjaszewski and N. V Tsarevsky, *Nat. Chem.*, 2009, **1**, 276–288.

71   LAMMPS Molecular Dynamics Simulator, *http://lammps.sandia.gov*.

72   G. Porod, in *Small Angle X-ray Scattering*, eds. O. Glatter and O. Kratky, Academic Press, New York, 1982.

73   A. Singh, A. Singh and A. Chakraborti, *J. Chem. Phys.*, 2017, **147**, 124902.

74   Y. Oono and S. Puri, *Phys. Rev. Lett.*, 1987, **58**, 836–839.

75   Y. Oono and S. Puri, *Phys. Rev. A*, 1988, **38**, 434–453.

76   S. Majumder and S. K. Das, *Phys. Chem. Chem. Phys.*, 2013, **15**, 13209.

77   D. Frenkel and B. Smit, *Understanding molecular simulation: from algorithms to applications*, Academic Press, 2001.

78   Y. Oono and S. Puri, *Mod. Phys. Lett. B*, 1988, **02**, 861–867.

**Figures and Captions**

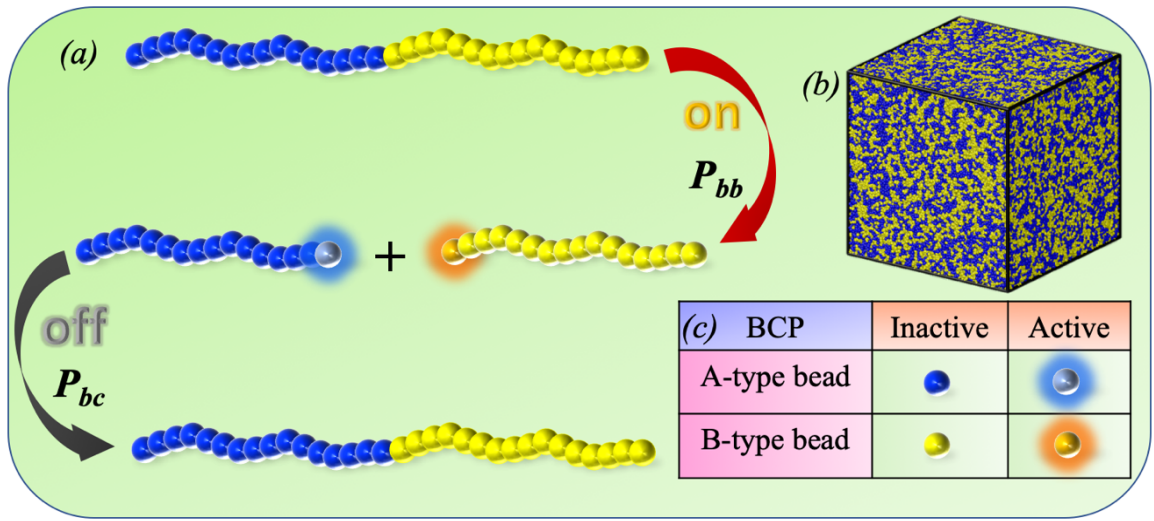

**Figure 1:** (a) Schematic of the bond breaking and bond formation reactions between the two blocks (depicted in the blue and yellow chains) in the presence ("on") and absence ("off") of the light, respectively. (b) An equilibrated BCP system (50% *A*-block and 50% *B*-block; critical composition) for $t = 2 \times 10^3$ simulation times. (c) The following colors mark the beads: inactive *A*-type (blue), active *A*-type radical (glowing cyan), inactive *B*-type (yellow), and active *B*-type radical (glowing orange). The bond breaking ($P_{bb}$; also represents light intensity) and bond formation ($P_{bc}$) probabilities can be varied independently.

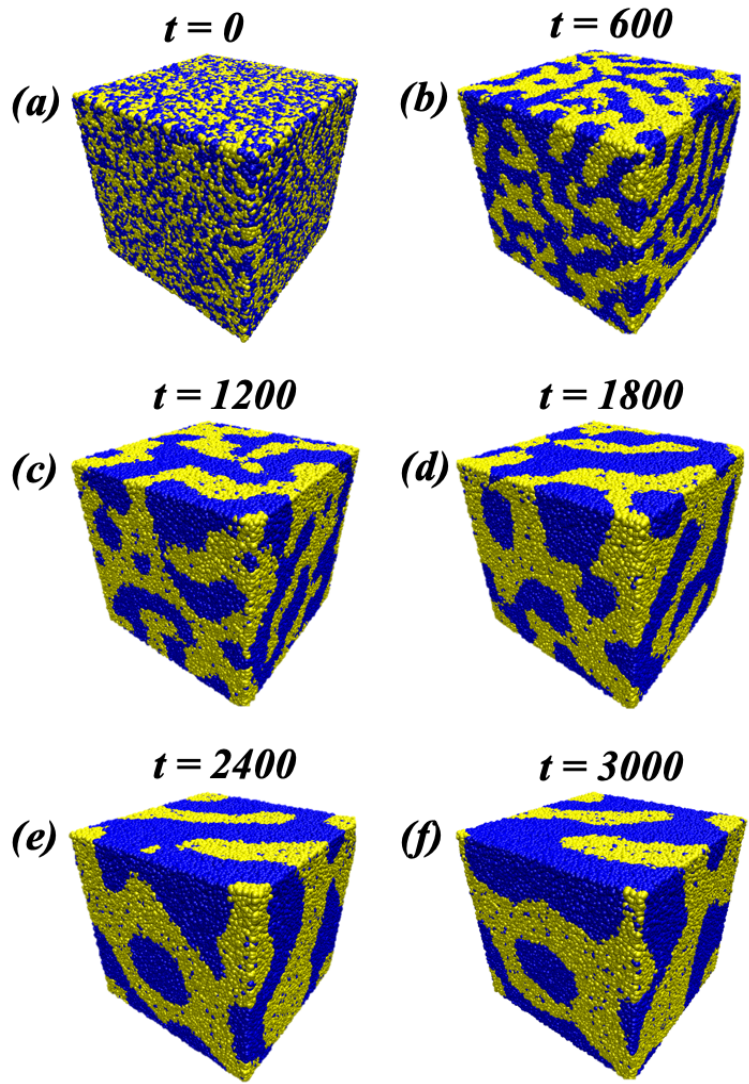

**Figure 2:** (a) Evolution morphology of BCP ($A_n B_n$; $n = 16$) melt after completing each cycle of $t = 600$ for set 1: (b) off → (c) on → (d) off → (e) on → (f) off when quenched from the initial homogeneously mixed state as shown in (a). The blue and yellow beads represent $A$-rich and $B$-rich domains, respectively.

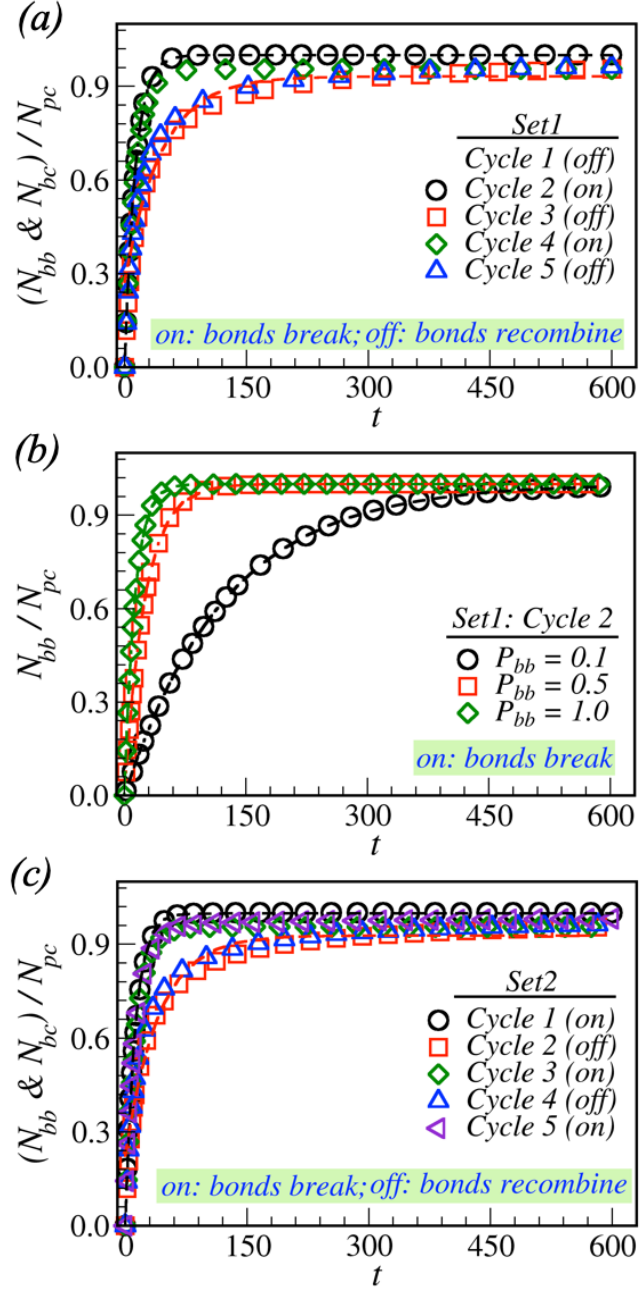

**Figure 3:** (a) Cumulative bonds broken, $N_{bb}(t)$ (in unit of total number of light sensitive bonds, $N_{pc}$ at $t = 0$) between two blocks under the light "on" cycles (2 & 4; $P_{bb} = 1.0$), and cumulative bonds recombined, $N_{bc}(t)$ in light "off" cycles (3, & 5; $P_{bc} = 1.0$) for set 1. (b) Change in $N_{bb}(t)$ at different values of $P_{bb}$ (shown in legends) for cycle 2 in set 1. (c) $N_{bb}(t)$ or $N_{bc}(t)$ during each period for set 2 (on → off → on → off → on). The dashed lines in each plot represent the exponential fits, $N_{bb}(t)/N_{pc} \sim \left(1 - e^{-\lambda P_{bb} t}\right)$, where $\lambda$ is a fitting parameter for the corresponding simulated data.

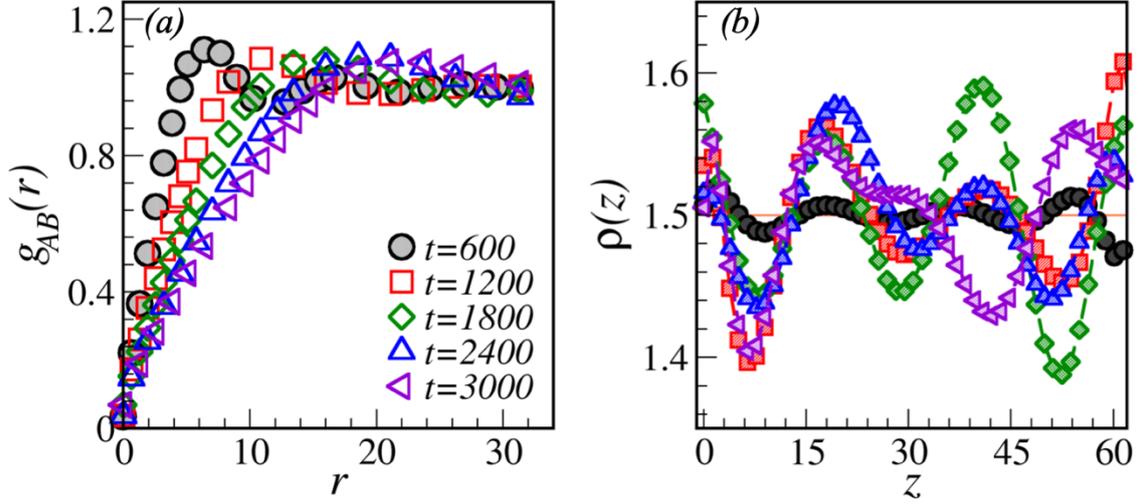

**Figure 4:** (a) Comparison of the radial distribution functions ($g_{AB}(r)$) of *A*-type beads around *B*-type at different times (depicted by the various symbol types) for the morphologies exhibited in Fig. 2a. (b) The average number density profile of *A*-type beads along the *z*-direction at different times as given in (a) with various symbol types. The averaging is done over five ensembles. Plots are done for the set 1 at $P_{bb} = 1.0$ and $P_{bc} = 1.0$, respectively.

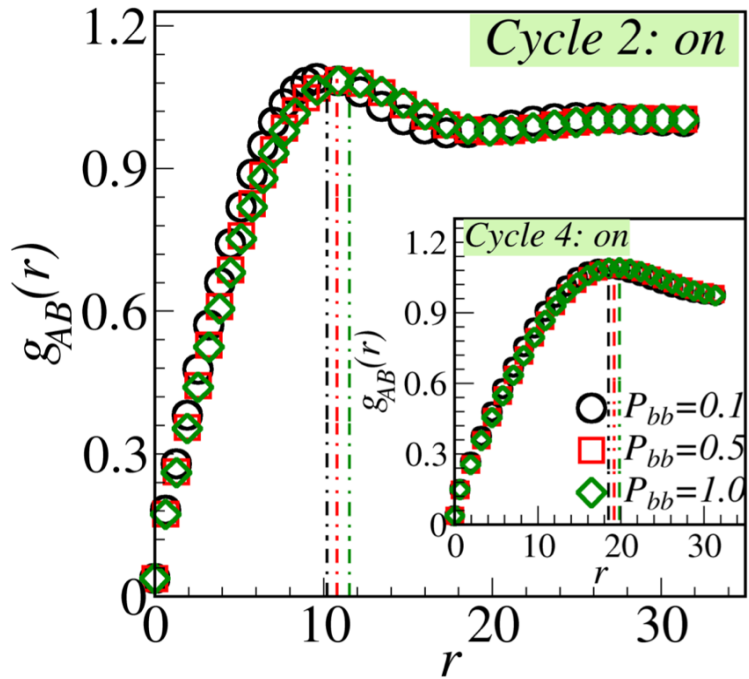

**Figure 5:** Comparison of the radial distribution functions ($g_{AB}(r)$) at different bond-breaking probabilities (denoted by different symbols) under illumination (cycle 2: on) for the set 1. The inset exhibits the same comparison for the fourth cycle.

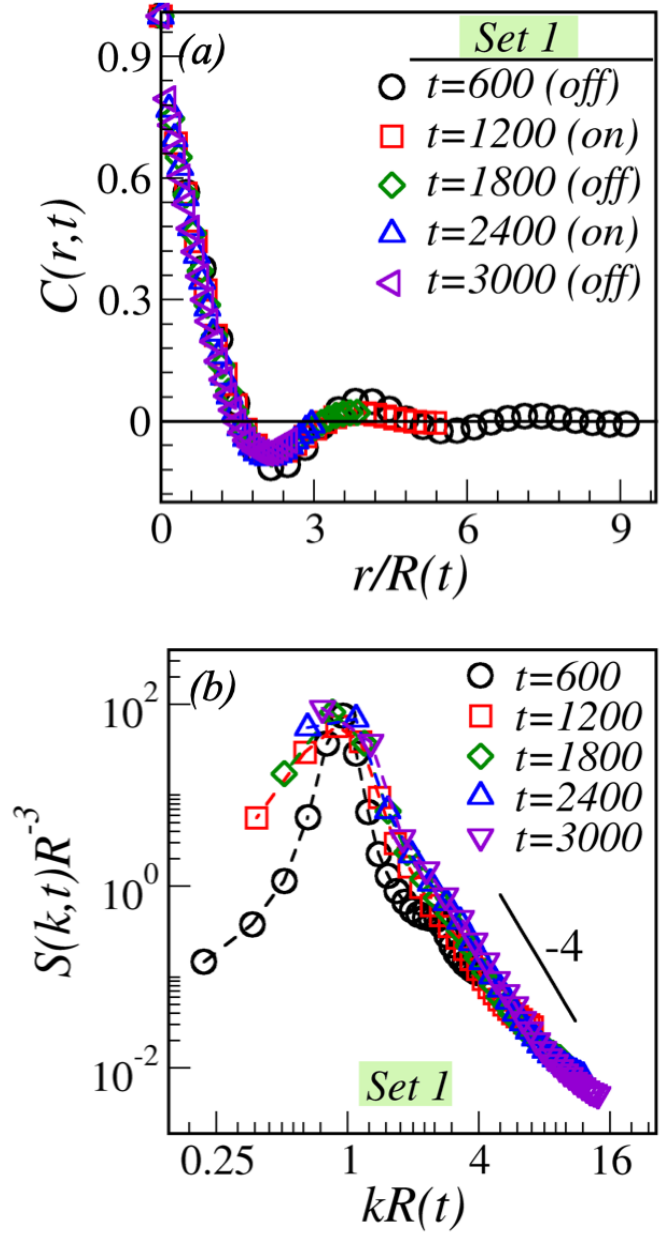

**Figure 6:** (a) The scaled correlation function, $C(r,t)$ versus $r/R(t)$ after completing each cycle for the set 1. The related evolution morphologies are shown in Fig. 2. The solid line displays the zero crossings of the correlation function. (b) Illustrates the scaled structure factor, $S(k,t)R(t)^{-3}$ versus $kR(t)$ plot for the same data sets as in (a) on a logarithmic scale. A solid line with slope -4 signifies the Porod's law, *i.e.*, $S(k,t) \sim k^{-4}$ for $k \to \infty$.

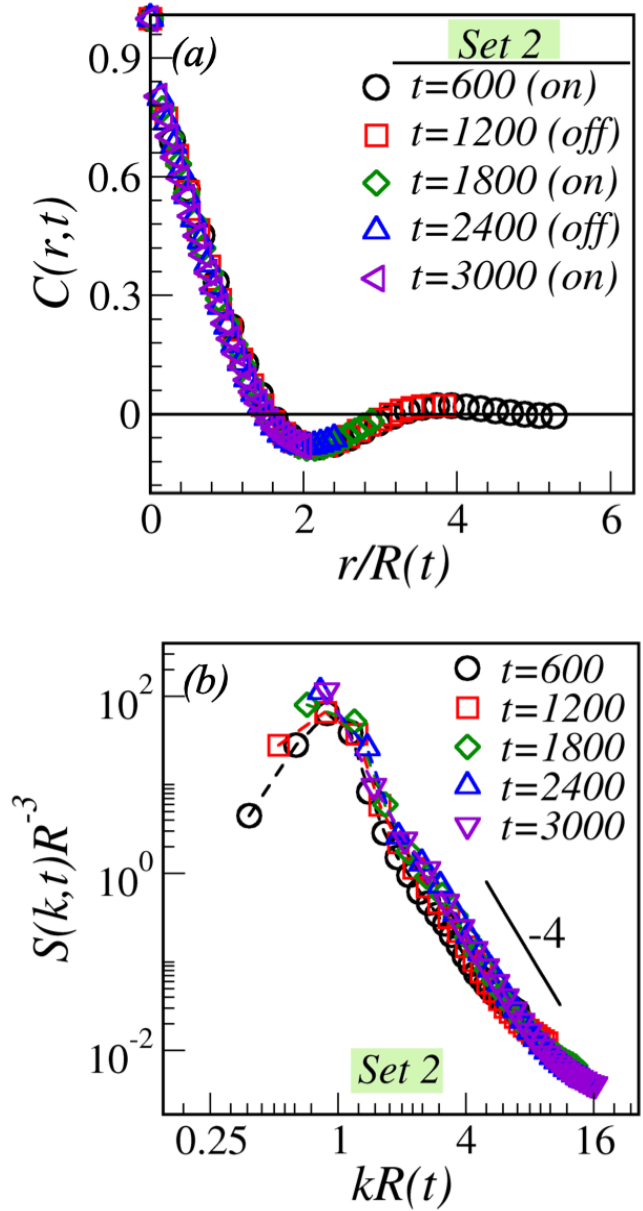

**Figure 7:** (a) Comparison of $C(r,t)$ versus $r/R(t)$ at the end of each cycle for the set 2 (on → off → on → off → on). (b) Explains the corresponding structure factor, $S(k,t)R(t)^{-3}$ versus $kR(t)$ plot with Porod tail behavior indicated by the solid line of slope $-4$.

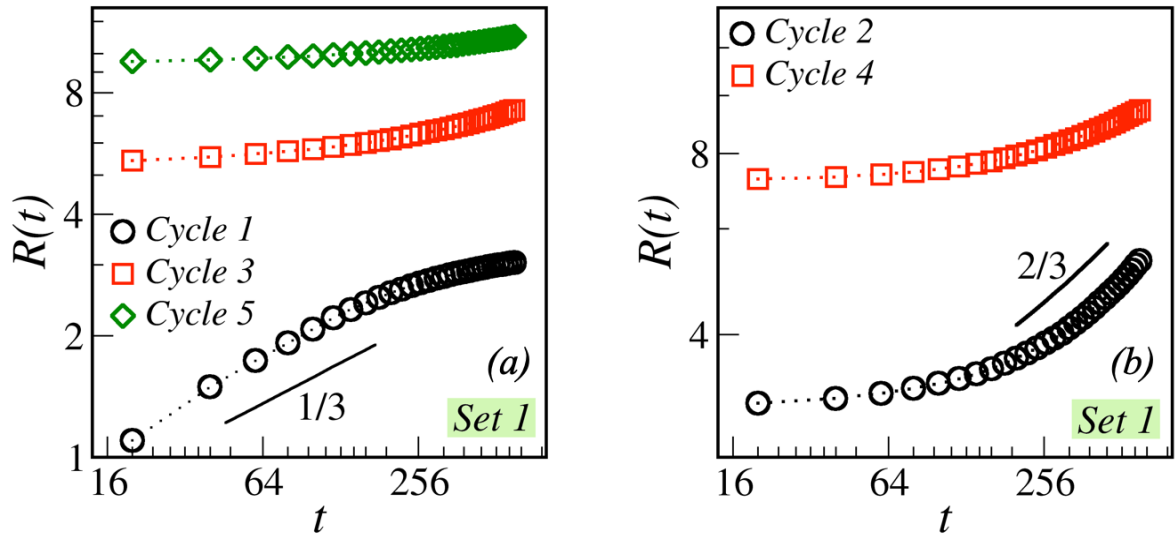

**Figure 8:** Logarithmic plot of the time dependence of the characteristic domain size, $R(t)$. (a) Exhibits the apparent microscopic domain growth during off cycles (1, 3 & 5) of set 1 when bonds combined between $A$ and $B$-blocks to form BCP blend. A solid line represents the expected diffusive growth exponent ($\phi \sim 1/3$) for a short period. (b) Shows the macroscopic domain growth during on cycles (2 & 4) when bonds between the blocks are broken, and the system behaves as a binary ($AB$) polymeric blend. A solid line of the slope, $\phi \sim 2/3$, corresponds to the expected inertial growth regimes for $d = 3$ fluids. Details of these plots are given in the text.

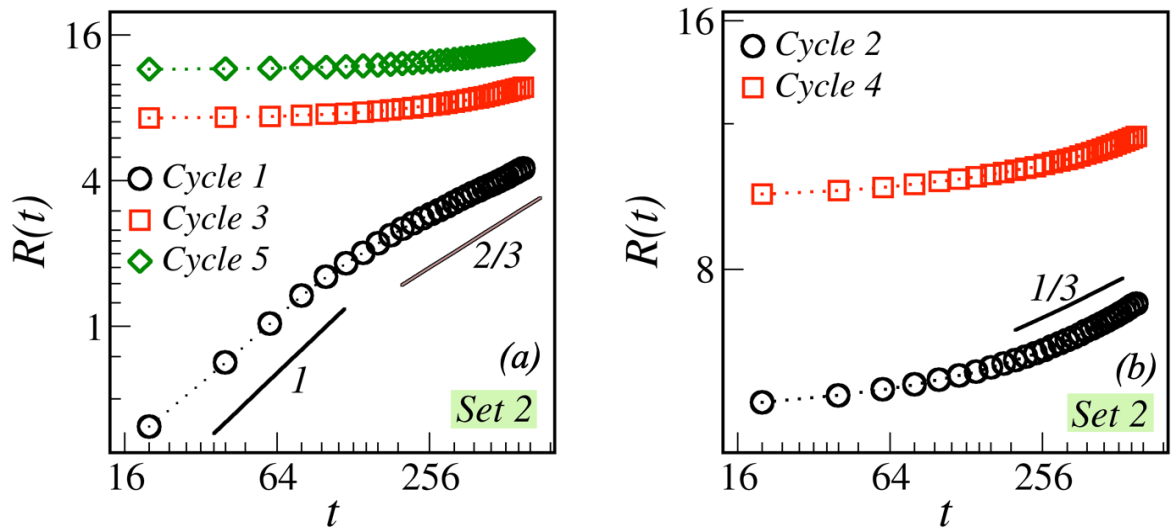

**Figure 9:** The time dependence of the average domain size, $R(t)$ on a logarithmic scale corresponding to the various cycles for set 2. (a) The solid lines of slope 1 and 2/3 represents the expected growth exponents for segregating binary ($AB$) blend in three-dimensions. (b) This plot exhibits the apparent microscopic domain growth during the off cycles (2 & 4) of BCP blend. A solid line represents the expected diffusive growth exponent ($\phi \sim 1/3$) for a short time.

**Photo-induced bond breaking during phase separation kinetics of block copolymer melts: A dissipative particle dynamics study**

Ashish Kumar Singh[1], Avinash Chauhan[1], Sanjay Puri[2*], and Awaneesh Singh[1*]

[1]Department of Physics, Indian Institute of Technology (BHU), Varanasi-221005, India.
[2]School of Physical Sciences, Jawaharlal Nehru University, New Delhi-110067, India.
*Emails: awaneesh.phy@iitbhu.ac.in; purijnu@gmail.com

**Supplementary Information**

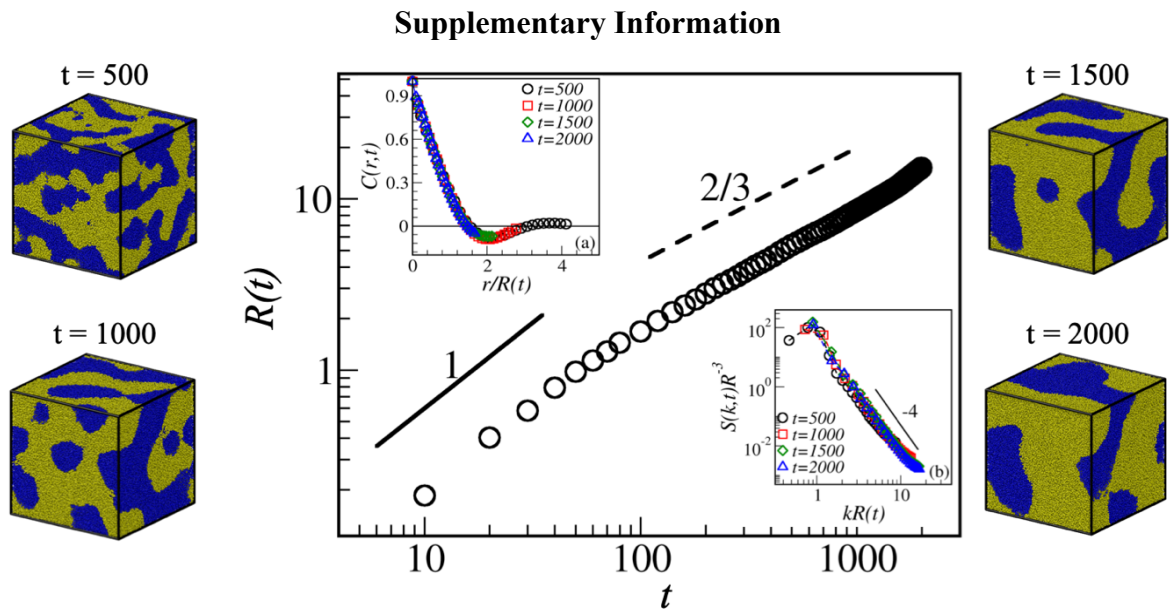

**Figure S1:** The evolution morphologies, scaling of its characterization functions, and the time dependence of average domain size are depicted for a binary ($AB$) polymer melt. Our model correctly illustrates the expected growth exponents in the viscous ($\phi = 1$) and inertial ($\phi = 2/3$) hydrodynamic regimes. The excellent data overlap of the correlation function and the structure factor at various times regard the presence of dynamical scaling (see the inset).

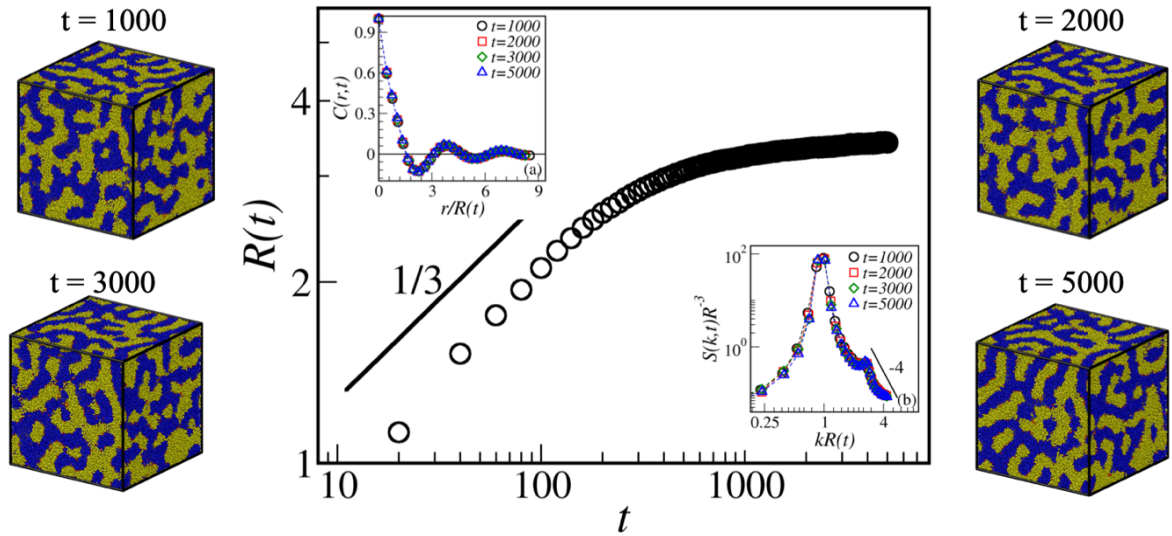

**Figure S2:** Shows the domain evolution, the scaling of its characterization functions, and the time dependence of the average domain size for BCP melt. Our model correctly illustrates the expected diffusive domain growth ($\phi = 1/3$) for a short while and then saturates to an average length scale. The data overlap of the characteristic functions at various times regard the presence of dynamical scaling (see the inset). The data oscillation around $C(r,t) = 0$, and a secondary peak in $S(k,t)$ confirms the formation of periodic domain structures in microphase separated BCP melt quenched below the critical temperature.

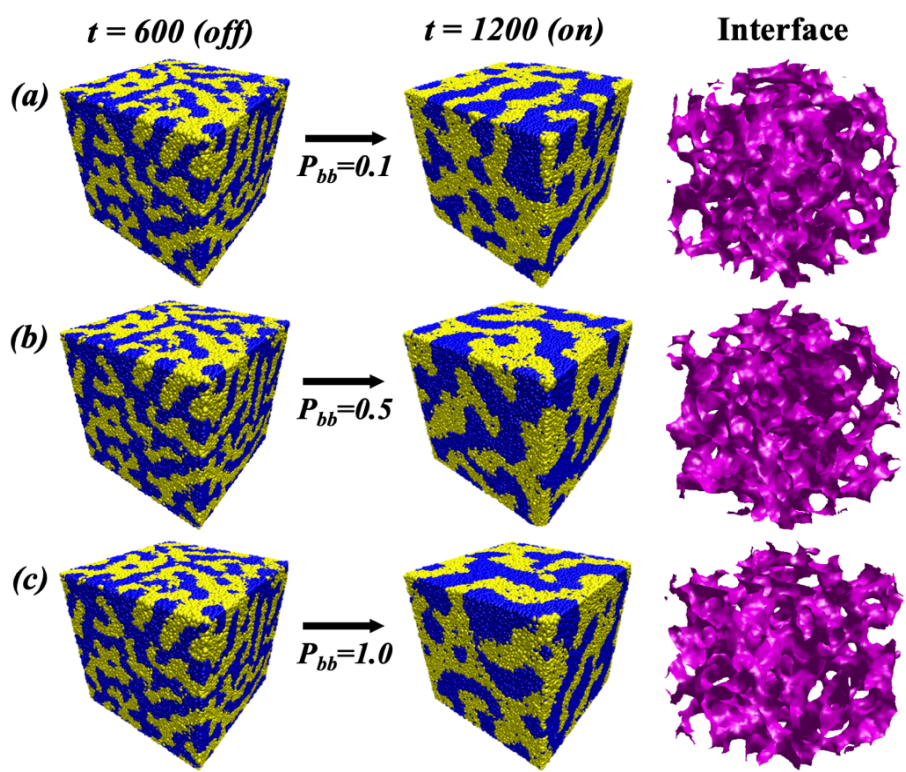

**Figure S3:** Evolution morphologies of BCP blend after completing off-cycle 1 and on-cycle 2 for set 1 at various bond-breaking probabilities: (a) $P_{bb} = 0.1$, (b) $P_{bb} = 0.5$, and (c) $P_{bb} = 1.0$. The first column shows microphase separated morphology at $t = 600$ as no bond breaking during cycle 1 (off). The second column displays macrophase separated morphology at $t = 1200$ as bonds break during cycle 2 (on) with different $P_{bb}$; therefore, dissimilar patterns. The third column illustrates the interface evolution between $A$ and $B$ phases for the morphologies shown in the second column.

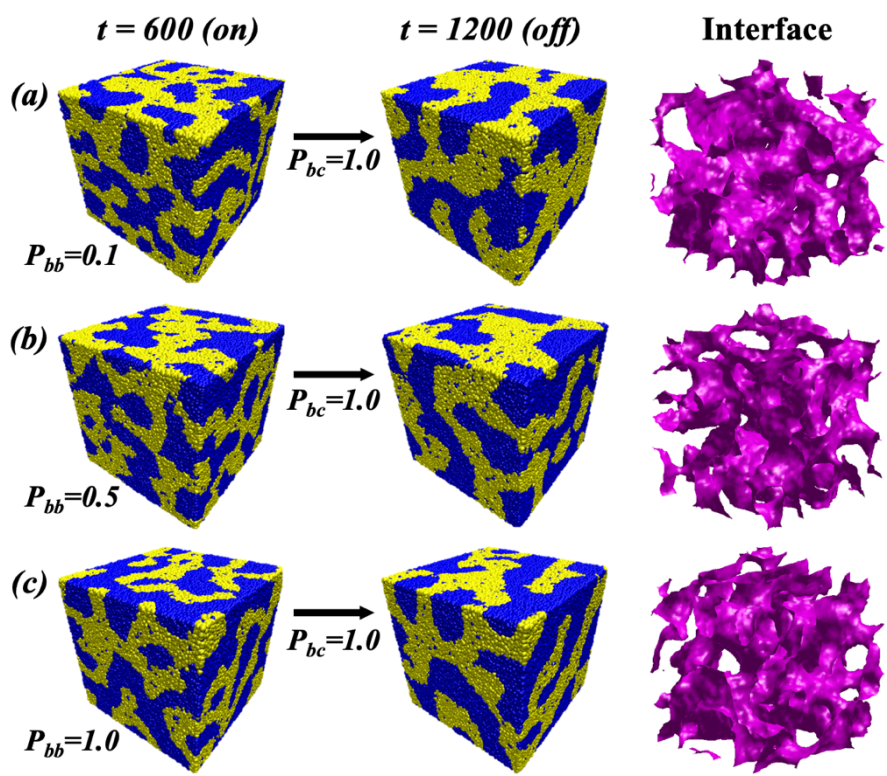

**Figure S4:** Evolution morphologies of BCP blend for set 2 at the end of on-cycle 1 with various bond-breaking probabilities: (a) $P_{bb} = 0.1$, (b) $P_{bb} = 0.5$, and (c) $P_{bb} = 1.0$., and at the end of off-cycle 2 with a bond combination probability, $P_{bc} = 1.0$. The first column shows macrophase separated morphology at $t = 600$ due to bond breaking through cycle 1 (on) with diverse $P_{bb}$; thus, varied patterns. The second column displays microphase separated morphologies at $t = 1200$ during cycle 2 (off) when bonds recombine with a probability, $P_{bc}$; different patterns as the domain formed during cycle 1 for each $P_{bb}$ is dissimilar. The third column illustrates the interfaces for morphologies shown in the second column.

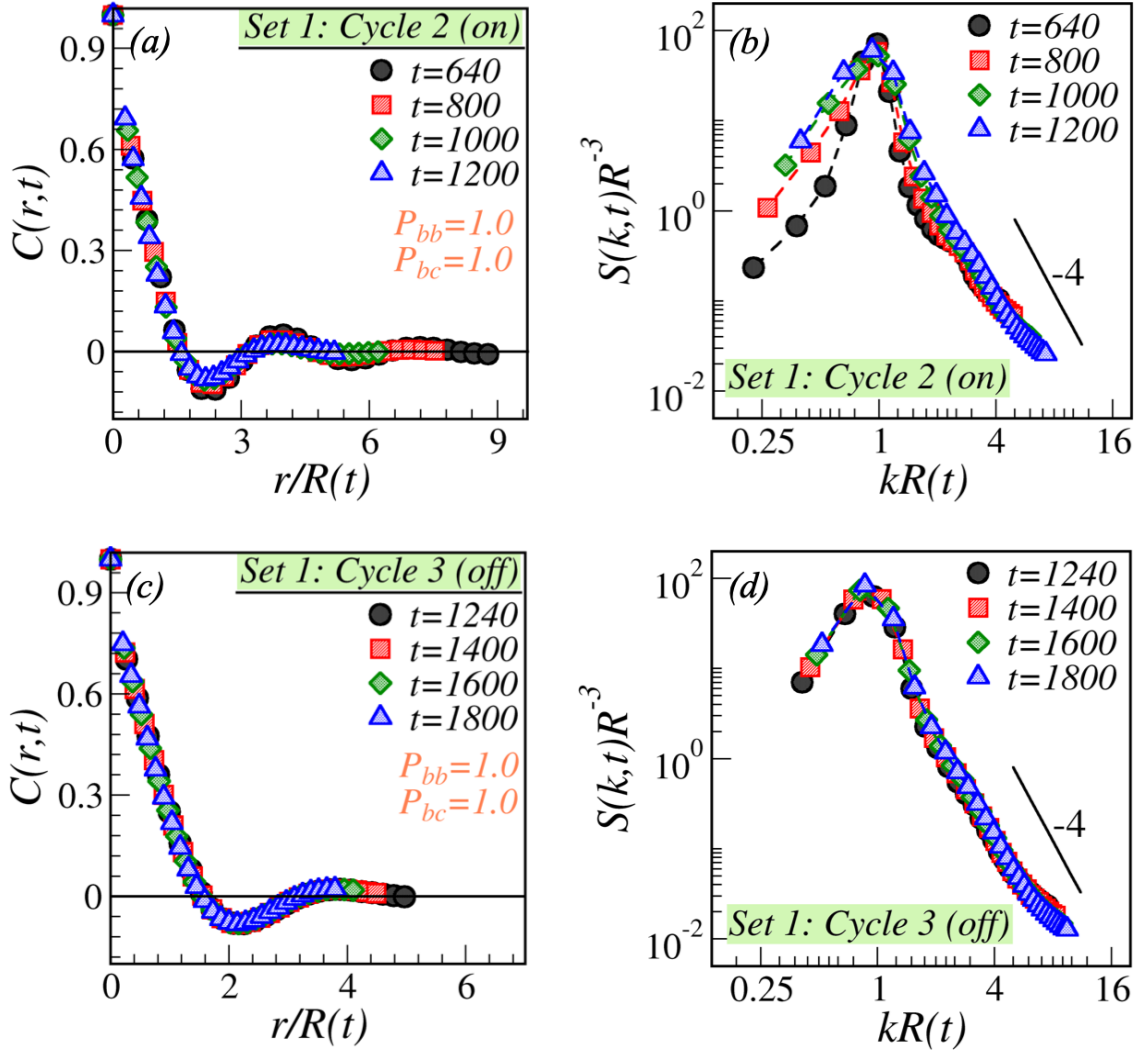

**Figure S5:** (a-b) Scaling of the correlation function ($C(r,t)$ vs. $r/R(t)$) and the structure factor ($S(k,t)R(t)^{-3}$ vs. $kR(t)$) for the evolution through cycle 2 (bond breaking on) for set 1. The system shows a small deviation from the scaling for early times (black and red curves at $t = 640$ and $t = 800$, respectively) when the kinetics is typically chemically controlled (bond breaking). However, we noted an excellent scaling in the system at late times (green and blue curves at $t = 1000$ and $t = 1200$, respectively). (c-d) We plot $C(r,t)$ vs. $r/R(t)$ and $S(k,t)R(t)^{-3}$ vs. $kR(t)$ for the evolution through cycle 3 (bonds are recombining) for the same set as in (a-b). The data overlap at various times reveals an excellent scaling in the system. The bond-breaking probability is set to $P_{bb} = 1.0$ during on cycles, and bond combination probability is set to $P_{bc} = 1.0$ during off periods. Each data set is averaged over five ensembles. A solid line with slope -4 signifies the Porod's law, *i.e.,* $S(k,t) \sim k^{-4}$ for $k \to \infty$.

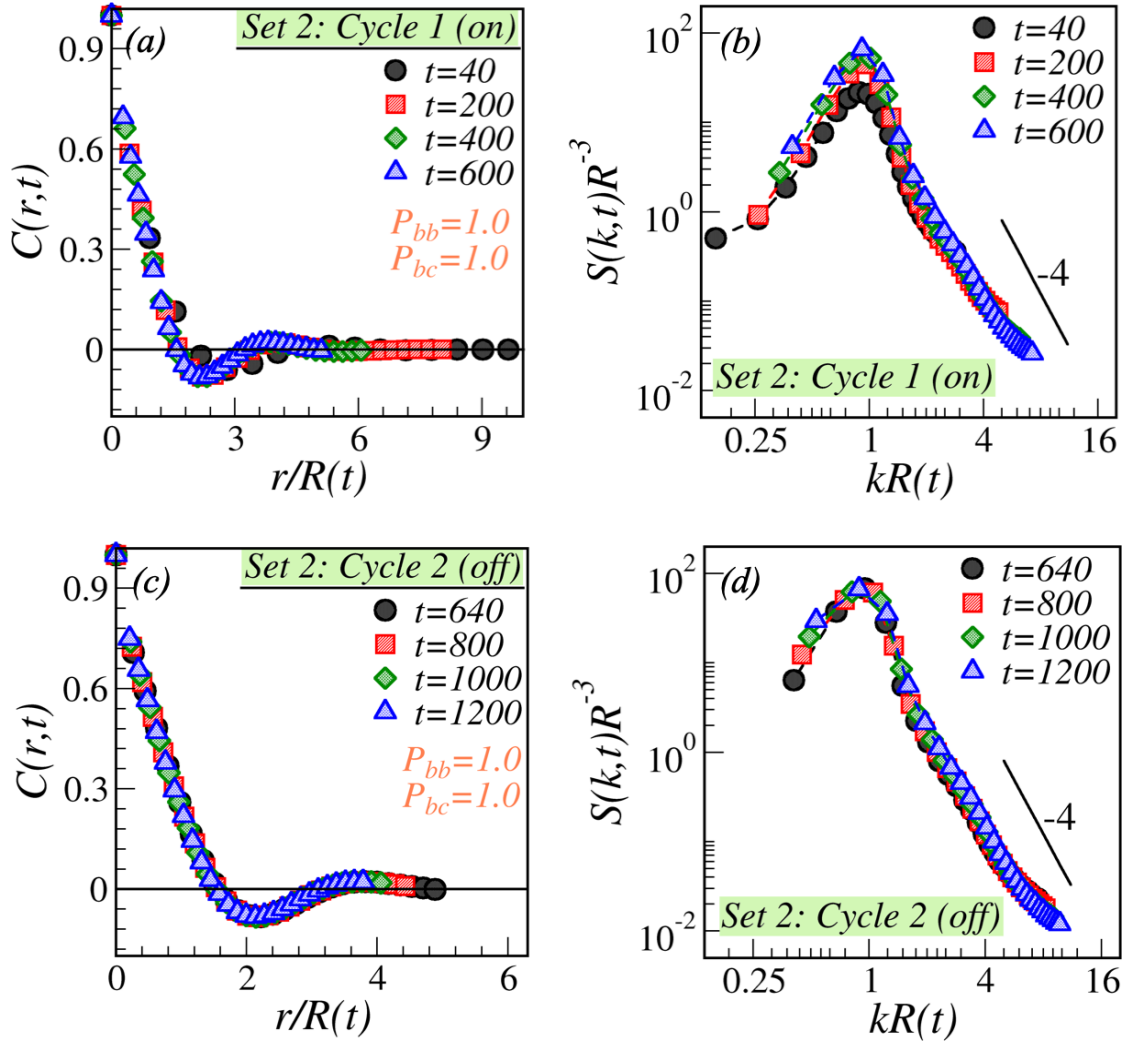

**Figure S6:** (a-b) We plot the correlation function ($C(r,t)$ vs. $r/R(t)$) and the structure factor ($S(k,t)R(t)^{-3}$ vs. $kR(t)$) for the evolution through cycle 1 (bond breaking on) for the set 2. Similar to the first on-cycle in set 1, a small deviation from the scaling is observed at early times (black and red curves at $t = 40$ and $t = 200$, respectively) for the chemically controlled regime. However, we note an excellent scaling at late times (green and blue curves at $t = 400$ and $t = 600$, respectively). (c-d) The plots $C(r,t)$ vs. $r/R(t)$ and $S(k,t)R(t)^{-3}$ vs. $kR(t)$ for the evolution through cycle 2 (bonds are recombining) for the same set as in (a-b) reveal an excellent scaling in the system. The bond-breaking probability is set to $P_{bb} = 1.0$ during on cycles, and bond combination probability is set to $P_{bc} = 1.0$ during off cycles. Each data set is averaged over five ensembles.

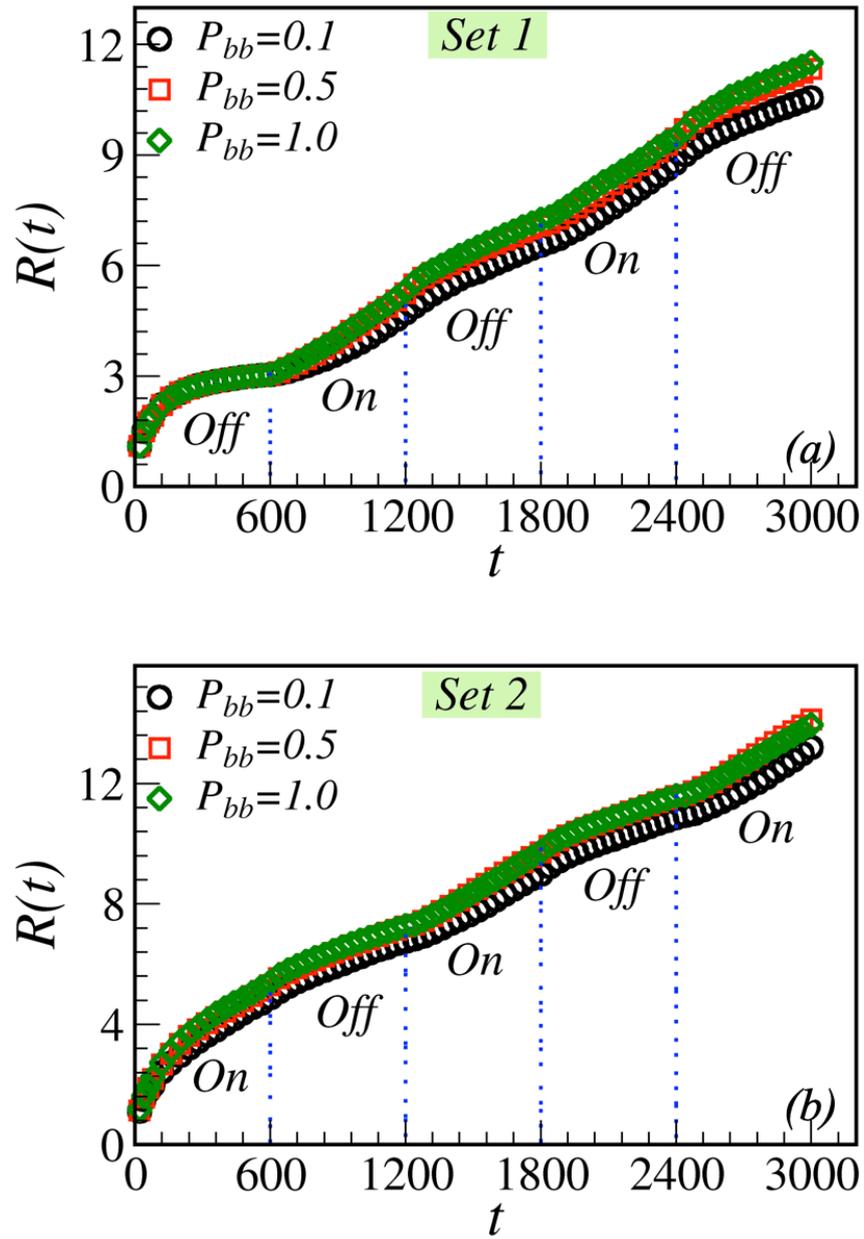

**Figure S7:** Time dependence of the average domain size for the set 1 is shown in (a) and for the set 2 is exhibited in (b) for the different cases represented by the various symbol types.